\documentclass[aps,prc,fleqn,floatfix,twocolumn,superscriptaddress,twoside]{revtex4}

\usepackage[dvipdfm]{graphicx}

\usepackage{amsmath,amssymb,amsfonts}

\usepackage{color}

\newcommand{\mean}[1]{\left\langle #1 \right\rangle} 

\begin{document}

\title{Scaling patterns of the suppression of $\pi^0$ yields in Au+Au collisions at \\ $\sqrt{s_{NN}}=200$~GeV:
       links to the transport properties of the QGP}
\author{ Roy~A.~Lacey} 
\author{ N.~N.~Ajitanand} 
\author{ J.~M.~Alexander}
\author{ X.~Gong}
\author{ J.~Jia}
\author{A.~Taranenko}
\author{Rui Wei}
\affiliation{Department of Chemistry, 
Stony Brook University, \\
Stony Brook, NY, 11794-3400, USA}


\date{\today}
\begin{abstract}

	Suppression measurements for neutral pions ($\pi^0$) are used to investigate 
the predicted path length ($L$) and transverse momentum ($p_T$) dependent jet quenching patterns of 
the hot QCD medium produced in Au+Au collisions at $\sqrt{s_{NN}}=200$\,GeV. 
The observed scaling patterns 
show the predicted trends for jet-medium interactions dominated by radiative 
energy loss. 
They also allow simple estimates of the transport coefficient $\hat{q}$ and the 
ratio of viscosity to entropy density $\eta/s$. These estimates indicate that the 
short mean free path ($\lambda$) in the QCD medium leading to hydrodynamic-like flow with a small 
value of $\eta/s$, is also responsible for the strong suppression observed.

\end{abstract}
\maketitle


One of the important discoveries at the Relativistic Heavy Ion Collider (RHIC) 
at Brookhaven National Laboratory (BNL), has been the observation that high-$p_T$ 
hadron yields are suppressed in central and mid-central A+A collisions when compared to 
the binary-scaled yields from p+p collisions  \cite{Adcox:2001jp}. 
This observation has been attributed to jet-quenching \cite{Gyulassy:1993hr} -- the 
process by which hard scattered partons interact and loose energy in the hot and 
dense quark gluon plasma (QGP) produced in the collisions. Subsequent to such interactions, 
the partons which do emerge, then fragment into topologically aligned hadrons (jets) 
which provide the basis for the $\pi^0$ suppression measurements.

	There is considerable current interest in   
the use of jet quenching as a quantitative tomographic probe of the QGP.  
Recent theoretical efforts have centered on investigations of the energy loss 
mechanism for scattered partons which propagate through this medium.
Two such mechanisms are; (i) scatterings off thermal
partons in binary elastic collisions and (ii) Gluon bremsstrahlung with the 
Landau-Pomeranchuk-Migdal (LPM) \cite{Migdal:1956tc} effect. The latter 
has been investigated via different formalisms \cite{Baier:1996kr, Kovner:2003zj, 
Zakharov:1996fv, Gyulassy:2000er, Dokshitzer:2001zm, Wang:2001ifa, Majumder:2007hx}. 
Studies of the relative importance of both jet quenching mechanisms 
have also been made \cite{Mustafa:2003vh, Adil:2006ei,
Zakharov:2007pj, Renk:2007id}. To date, a conclusive mechanistic 
picture has not yet emerged. 

	Initial quantitative studies with models which incorporate the time evolution of the 
QGP medium via relativistic ideal (3+1)-dimensional hydrodynamical simulations, are currently 
underway \cite{Qin:2007rn,Bass:2008rv}. 
However, the value of such studies rests heavily on accurate knowledge of the 
dominant mechanism/s for jet quenching. Therefore, it is important to pursue 
validation tests which can lend insight or provide a clear distinction between different 
energy loss mechanisms.

\begin{figure}[t]
\includegraphics[width=1.0\linewidth]{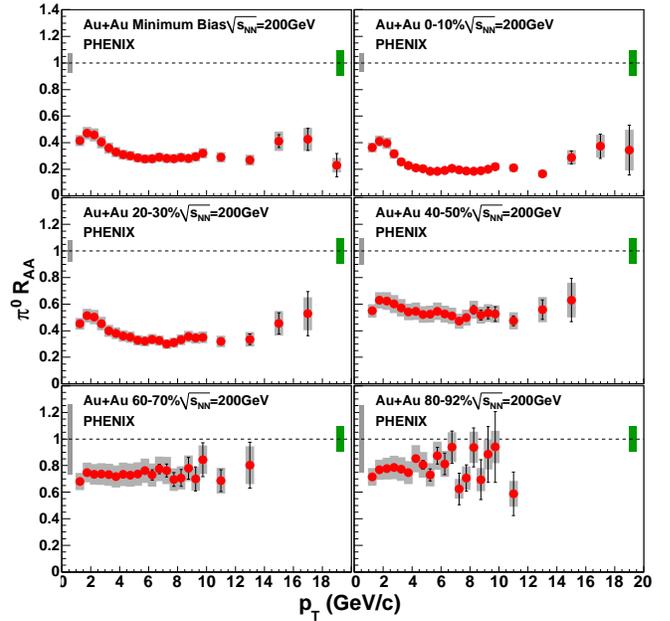}
\caption{\label{phenixRAA} (Color online) Nuclear modification factor ($R_{\rm AA}$) for
$\pi^0$s reproduced from Ref.~\cite{Adare:2008qa}.  Error bars
indicate statistical and $p_{\rm T}$-uncorrelated errors; boxes indicate $p_{\rm T}$-correlated 
errors.  The single box around $R_{\rm AA}$=1 on 
the left is the error due to $N_{\rm coll}$; the single box on the right is the
overall normalization error of the p+p reference spectrum.
}
\end{figure}

	The experimental probe commonly exploited for jet-quenching studies 
in AA collisions is the nuclear modification factor ($R_{\rm AA}$);
\[
   R_{\rm AA}(p_T) = \frac{1/{N_{\rm evt}} dN/dydp_{\rm T}}{\mean{T_{\rm AA}} d\sigma_{pp}/dydp_{\rm T}}, 
\]
where $\sigma_{pp}$ is the particle production cross section in p+p collisions 
and $\mean{T_{\rm AA}}$ is the nuclear thickness function
averaged over the impact parameter range associated with a given 
centrality selection
\[
\langle T_{AA}\rangle\equiv
\frac {\int T_{AA}(\mathbf{b})\, d\mathbf{b} }{\int (1- e^{-\sigma_{pp}^{inel}\, T_{AA}(\mathbf{b})})\, d\mathbf{b}}.
\]
The corresponding average number of nucleon-nucleon collisions, 
$\langle N_{coll}\rangle=\sigma_{pp}^{inel} \langle T_{AA}\rangle$,
is routinely obtained via a Monte-Carlo Glauber-based 
model calculation \cite{Miller:2007ri,Alver:2006wh}.

	In this letter we use $R_{\rm AA}$ measurements to perform validation tests which 
addresses the question of whether or not medium induced gluon radiation is a dominant 
mechanism for jet-energy loss.

	The measurements employed for these tests are the recently published $R_{\rm AA}(p_T)$
and $R_{\rm AA}(\Delta\phi,p_T)$ data for $\pi^0$ \cite{Adare:2008qa,Adler:2006bw}. 
Here, $\Delta\phi$ is the azimuthal angle relative to the reaction plane. 
A subset of these data is shown as a function of $p_T$ 
for several centrality selections in Fig.~\ref{phenixRAA}. The minimum bias results
in the top left panel indicate that, for $2.5 \alt p_T  \alt 5$ GeV/c,
suppression increases with $p_T$ to the value $R_{\rm AA} \sim$0.3. 
By contrast, a much weaker $p_T$ dependence is observed for $p_T  \agt 5$ GeV/c, with a hint 
that the suppression decreases as $p_T$ increases. The same trends are evident for all 
centrality selections spanning central and mid-central collisions, albeit with different 
absolute magnitudes. These trends provide an important constraint 
for our tests, as discussed below.

	To perform validation tests on the data for medium induced gluon radiation, we use the 
``pocket formula'' of Dokshitzer and Kharzeev (DK)  \cite{Dokshitzer:2001zm}. 
This formula gives the quenching of the 
transverse momentum spectrum for jets produced from scattered light 
partons \cite{Dokshitzer:2001zm} as;
\begin{eqnarray}
R_{\rm AA}(p_T,L) \simeq \exp \left[- {2 \alpha_s C_F \over \sqrt{\pi}}\ 
L\,\sqrt{\hat{q}\frac{{\cal{L}}}{p_T}}\, \right] \nonumber \\
{\cal{L}} \equiv \frac{d}{d\ln p_T} 
\ln \left[ {d \sigma_{pp} \over d p_{T}^2}( p_{T})\right], 
\label{eq:DK1}	
\end{eqnarray}
where $\alpha_s$ is the strong interaction coupling strength, $C_F$ is the color factor,
$L$ is the path length [of the medium] that the parton traverses and $\hat{q}$ is 
the transport coefficient which reflects the squared average transverse 
momentum exchange per unit path length, between the medium and the parton. 

\begin{figure}[t]
\includegraphics[width=1.0\linewidth]{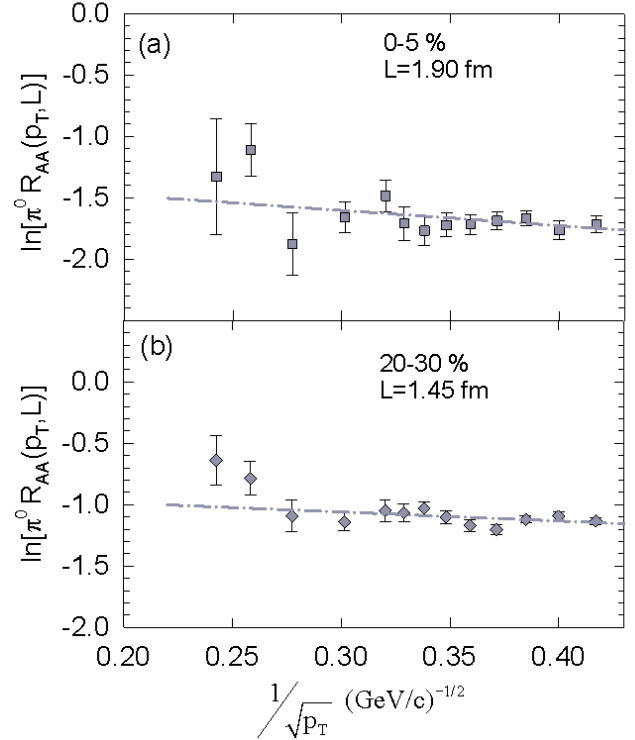}
\caption{(Color online) $\ln\left[R_{\rm AA}(p_T,L)\right]$ vs. $1/\sqrt(p_T)$ ($p_T \agt 5$~GeV/c) for centrality selections 
of 0-5\% (a) and 20-30\% (b). Error bars are statistical only. 
$p_{\rm T}$-correlated systematic errors are about 12\% and the systematic 
errors associated with $N_{\rm coll}$ and the overall normalization of the p+p 
reference spectrum are $\sim 7$\% and 10\% respectively.
The dot-dashed curve in each panel is a fit to the data (see text).
} 
\label{Fig3_pT}
\end{figure}

	Here, there are two essential points. First, Eq.~\ref{eq:DK1} is derived with an explicit 
assumption that the mechanism for energy loss is medium induced gluon radiation \cite{collisional}.
Second, this equation gives specific testable predictions for the dependence of $R_{\rm AA}(p_T,L)$ 
on $L$ and $p_T$ for $p_T > 5$~GeV/c. 
That is, $\ln\left[R_{\rm AA}(p_T,L)\right]$ should scale as $L$ and $1/\sqrt(p_T)$ respectively
(ie. $\ln\left[R_{\rm AA}(p_T,L)\right]$ should show a linear dependence on both $L$ and $1/\sqrt(p_T)$
if medium induced gluon radiation is indeed the dominant energy loss mechanism).

	 To study the $L$ dependence, the transverse size of 
the system $\bar{R}$ was used as an estimate for the angle averaged 
path length $L$, as well as to determine its value in- ($L_{x}$) and out- ($L_{y}$) 
of the reaction plane. For each centrality selection, the number of participant nucleons 
$N_{\rm part}$, was estimated  via a Monte-Carlo Glauber-based 
model \cite{Miller:2007ri,Alver:2006wh}. The corresponding 
transverse size $\bar{R}$ was then determined from the distribution of these nucleons in the 
transverse ($x,y$) plane via the same Monte-Carlo Glauber model:
\[
\frac{1}{\bar{R}} = \sqrt{\left(\frac{1}{\sigma_x^2}+\frac{1}{\sigma_y^2}\right)},
%
%
%
\]
%
where $\sigma_x$ and $\sigma_y$ are the respective root-mean-square widths of
the density distributions; 
here, averaging is performed over configurations. 
For these calculations, the initial entropy profile in the transverse
plane was assumed to be proportional to a linear combination
of the number density of participants and binary collisions \cite{Hirano:2009ah,Lacey:2009xx}.
The latter assures that the entropy density weighting is constrained by hadron multiplicity 
measurements.

\begin{figure}[htb]
\includegraphics[width=0.95\linewidth]{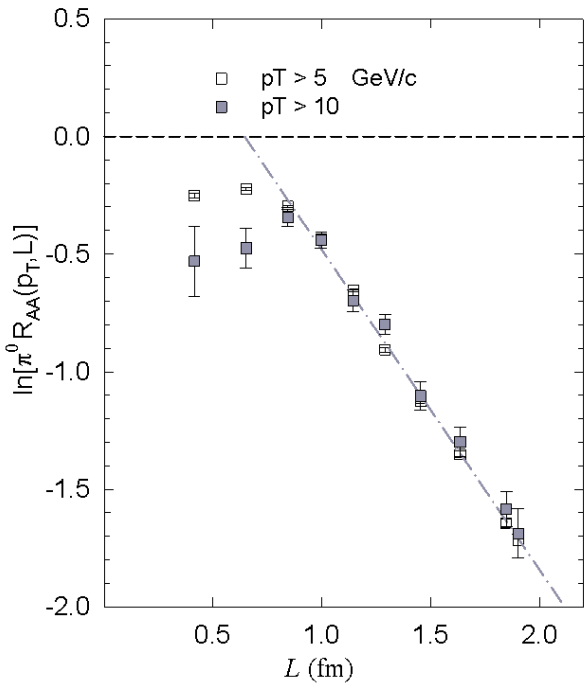}
\caption{ (Color online) $\ln\left[R_{\rm AA}(p_T,L)\right]$ vs. $L$ for two $p_T$ selections as indicated.
         Error bars are statistical only. The systematic error associated with $N_{\rm coll}$ 
         ranges from $\sim 7$\% in central collisions to $\sim 30$\% in peripheral collisions; 
         the systematic error resulting from overall normalization of the p+p reference 
         spectrum is about 10\% (cf. Fig.\ref{phenixRAA}). The dot-dashed curve is a linear fit to the data 
         set for $p_T > 5$~GeV/c (see text).
} 
\label{Fig2_L}
\end{figure}

\begin{figure}[t]
\includegraphics[width=1.0\linewidth]{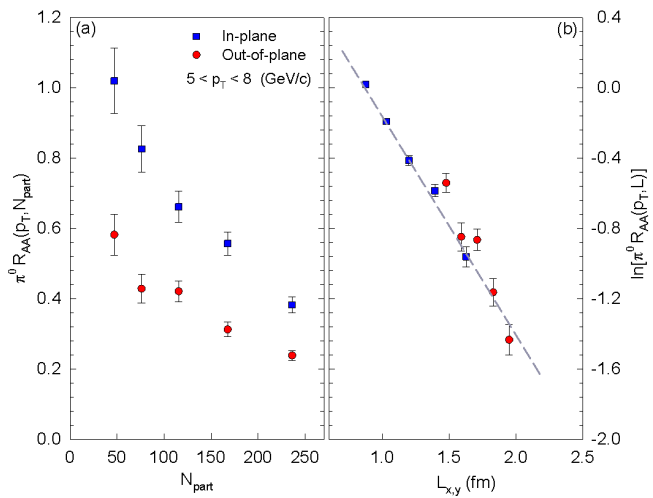}
\caption{(Color online) (a) $R_{\rm AA}(p_T,N_{\rm part})$ vs. $N_{\rm part}$ for $\pi^0$'s 
detected in- and out of the reaction plane. 
The data are taken from Ref.~\cite{Adler:2006bw} for the selection $5 < p_T < 8$ GeV/c. 
Error bars are statistical only. 
(b) $\ln\left[R_{\rm AA}(p_T,L)\right]$ vs. $L_{\rm x,y}$, the path length in- and out of 
the reaction plane. The systematic errors for both (a) and (b) are similar to those 
of Fig. \ref{Fig2_L}. The dashed curve is a linear fit to the data.
} 
\label{Fig4_Rxn}
\end{figure}

		The results from our validation tests for medium-induced gluon radiation 
are summarized in Figs.~\ref{Fig3_pT}~--~\ref{Fig4_Rxn}.
Figs. \ref{Fig3_pT} (a) and (b) show plots of $\ln\left[R_{\rm AA}(p_T,L)\right]$ vs. $1/\sqrt(p_T)$ for 
$p_T \agt 5$~GeV/c and centrality selections of 0-5\% (L=1.90 fm) and 20-30\% (L=1.45 fm) respectively. 
The dot-dashed curves in these figures represent a linear fit to the data. They show that, although
there is some scatter, the trend of the data is compatible with the $1/\sqrt(p_T)$ dependence 
predicted by Eq.~\ref{eq:DK1}. It is worth mentioning here that a similar $1/\sqrt(p_T)$ dependence is 
observed for the range $2.5 \alt p_T \alt 5$~GeV/c, but with a different (positive) slope.
%

Figure \ref{Fig2_L} shows the $L$ dependence of $\ln\left[R_{\rm AA}(p_T,L)\right]$ for two different $p_T$ 
selections as indicated. The dot-dashed curve shows a linear fit to the data for $p_T > 5$~GeV/c
and $L \agt 0.65$~fm, ie. the two data points corresponding to the two most peripheral collision 
centralities were excluded from the fit (note the large $N_{\rm coll}$ systematic errors for 
peripheral collisions). Fig.~\ref{Fig2_L} shows that the data trends  
for both $p_T$ selections ($p_T>5$ and $p_T>10$ GeV/c), validates the  
linear dependence on $L$ predicted by Eq.~\ref{eq:DK1}. The efficacy of this scaling 
pattern is not significantly influenced by systematic errors.

	Further evidence for this linear dependence is also evident in Fig.~\ref{Fig4_Rxn}
where we show results for $R_{\rm AA}$ measurements in- and out of the reaction 
plane for $\pi^0$s with $5 < p_T <8$ GeV/c \cite{Adler:2006bw}.  Panel (a) shows that,
for the same number of participants, $\pi^0$ suppression is larger out-of-plane 
than it is in-plane. However, panel (b) shows that, when plotted as 
$\ln\left[R_{\rm AA}(p_T,L_{x,y})\right]$ vs. $L_{x}$
and $L_{y}$, the in-plane and out-of-plane data show a single linear 
dependence on $L_{\rm x,y}$. This same dependence is found for all other 
reaction plane orientations for $p_T \agt 5$ GeV/c. This confirms that the azimuthal 
anisotropy of particle yields (at high $p_T$) stem from the path-length dependence of jet 
quenching. The dashed curve (fit to the data) in the figure reinforces the 
predicted linear dependence of these data on the path length (cf. Eq.~\ref{eq:DK1}).

	The fits to the data in Fig.~\ref{Fig2_L} (for the two $p_T$ cuts) indicate 
an intercept $L \approx 0.6$~fm, which suggests a minimum path length requirement 
for the initiation of $\pi^0$ suppression. Such a requirement is akin to the 
plasma formation or cooling times proposed in 
Refs.~\cite{Pantuev:2005jt,Liao:2005hj}. 

The corresponding slopes (for $p_T > 5$ and $p_T > 10$ GeV/c) indicate similar 
magnitudes with a hint of a small decrease in slope with increasing 
$\left\langle p_T\right\rangle$. 
This relatively weak $p_T$ dependence (which is also reflected in Fig.~\ref{Fig3_pT}), could be 
an indication that $\hat{q}$ has a dependence on $p_T$, 
eg. $\hat{q}=\hat{q_0}\left(\frac{p_T}{p_T^0}\right)^{\lambda}$
where $\lambda$ is a small fractional power \cite{CasalderreySolana:2007sw}.

	To obtain an estimate of the magnitude of $\hat{q}$, we use Eq.~\ref{eq:DK1} 
in conjunction with the slope, $-1.26 \pm 0.06$ fm$^{-1}$, extracted from 
Fig.~\ref{Fig2_L} for $p_T >10$ GeV/c. This gives the value $\hat{q} \approx 0.75$~GeV$^2$/fm
for the values $\alpha_s = 0.3$ \cite{Bass:2008rv}, 
$C_F = 9/4$ \cite{cfactor,Dokshitzer:2001zm} and ${\cal{L}} = n = 8.1 \pm 0.05$ \cite{Adler:2006bw}. 
This estimate of $\hat{q}$, which can be interpreted as a space-time average, 
is comparable to the recent estimates of $\sim 1 - 2$~GeV$^2$/fm
obtained from fits to hadron suppression data (for the most
central Au+Au collisions) within the framework of 
the higher twist (HT) expansion \cite{Zhang:2007ja,Majumder:2007ae} and the Gyulassy-Levai-Vitev (GLV) 
scheme \cite{Gyulassy:2000fs,Majumder:2007iu}.
It is, however, much less than the value recently extracted 
via the approach of Arnold, Moore and Yaffe (AMY) \cite{Bass:2008rv,Qin:2007zzf} and that of 
Armesto Salgado and Wiedemann (ASW) \cite{Bass:2008rv,Wiedemann:2000ez}.

The ratio of the shear viscosity ($\eta$) to the entropy density ($s$) can also 
be estimated as \cite{Majumder:2007zh};
\[
\frac{\eta}{s} \approx 1.25 \frac{T^3}{\hat{q}}
\]
where $T$ is the temperature. This estimate can be compared to the 
value of $4\pi\frac{\eta}{s} \approx 1.3 \pm 0.3$ extracted from 
flow data \cite{Lacey:2009xx} for $T \sim 220$~MeV \cite{Adare:2008fqa}. 
It is noteworthy that if we use this same temperature in conjunction with 
our extracted value for $\hat{q}$ (in the above equation), we obtain a strikingly 
similar estimate for $\eta/s$. We conclude therefore, that the relatively 
short mean free path in the plasma \cite{Lacey:2009xx} which drives  
hydrodynamic-like flow with small shear viscosity, is also responsible for 
the strong jet quenching observed.

	In summary, we have performed validation tests of the scaling properties of jet quenching 
to investigate the dominant mechanism for jet-energy loss. Our tests confirm 
the weak $1/\sqrt(p_T)$ dependence, as well as the linear dependence on 
path length ($L$ and $L_{x,y}$) predicted by Dokshitzer and Kharzeev, for jet suppression dominated 
by the mechanism of medium-induced gluon radiation in a hot and dense QGP. 
The quenching patterns indicate a minimum path length requirement 
for the initiation of $\pi^0$ suppression (corona), suggests a possible 
$p_T$ dependence for $\hat{q}$ and gives the estimate $\hat{q} \sim 1$~GeV$^2$/fm  which is 
comparable to that obtained within the framework of the HT and GLV models. This estimate also 
indicates a small value of $\eta/s$ which is in good agreement with that 
obtained via flow measurements.
The present study will have to 
be extended to the heavy quark systems to arrive at an even more definitive conclusion about 
the role of a perturbative radiative energy loss mechanism. Nonetheless, these results 
will undoubtedly provide important model constraints in future attempts to use jet-quenching 
as a quantitative tomographic probe of the QGP.

\section*{Acknowledgments}
We thank D.E. Kharzeev for crucial and interesting discussions.
This work was supported by the US DOE under contract DE-FG02-87ER40331.A008.
 

\bibliography{QhatRefs}

\begin{thebibliography}{35}
\bibitem[{cfa()}]{cfactor}
\bibinfo{note}{Here, it is assumed that $\pi^0s$ are dominantly produced via
  the fragmentation of gluons.}
\end{thebibliography}
\end{document}